\documentclass[letter]{aa} 
\usepackage{graphicx}
\usepackage[varg]{txfonts}
\begin{document}
   \title{Diffuse interstellar bands in fullerene planetary nebulae: the
   fullerenes - diffuse interstellar bands connection}

   \titlerunning{DIBs in Fullerene PNe}

   \author{D. A. Garc\'{\i}a-Hern\'andez\inst{1,2} \and J. J.  D\'{\i}az-Luis\inst{1,2} }
	  
\authorrunning{Garc\'{\i}a-Hern\'andez \&D\'{\i}az-Luis}

   \institute{Instituto de Astrof\'{\i}sica de Canarias, C/ Via L\'actea s/n,
   E$-$38200 La Laguna, Spain\\
              \email{agarcia@iac.es}
         \and
              Departamento de Astrof\'{\i}sica, Universidad de La Laguna (ULL),
	     E$-$38206 La Laguna, Spain\\
             }

   \date{Received December 14, 2012; accepted January 1, 2013}

 
\abstract
{We present high-resolution (R$\sim$15,000) VLT/UVES optical spectra of two
planetary nebulae (PNe; Tc 1 and M 1-20) where C$_{60}$ (and C$_{70}$)
fullerenes have already been found. These spectra are of high-quality (S/N $>$
300) for PN Tc 1, which permits us to search for the expected electronic
transitions of neutral C$_{60}$ and diffuse interstellar bands (DIBs).
Surprisingly, we report the non-detection of the most intense optical bands of 
C$_{60}$ in Tc 1, although this could be explained by the low C$_{60}$ column
density estimated from the C$_{60}$ infrared bands if the C$_{60}$ emission
peaks far away from the central star. The strongest and most common DIBs in both
fullerene PNe are normal for their reddening. Interestingly, the very broad 4428
\AA\ DIB and the weaker 6309 \AA~DIB are found to be unusually intense in Tc 1.
We also report the detection of a new broad (FWHM$\sim$5 \AA) unidentified band
at $\sim$6525 \AA. We propose that the 4428\AA\ DIB (probably also the 6309\AA\
DIB and the new 6525 \AA\ band) may be related to the presence of larger
fullerenes (e.g., C$_{80}$, C$_{240}$, C$_{320}$, and C$_{540}$) and buckyonions
(multishell fullerenes such as C$_{60}$@C$_{240}$ and
C$_{60}$@C$_{240}$@C$_{540}$) in the circumstellar envelope of Tc 1.}


   \maketitle
%

\section{Introduction}

The diffuse interstellar bands (DIBs) have remained a mystery for astronomers
since their discovery about ninety years ago (Heger 1922); they are one of the
long-standing problems in the interstellar medium (ISM). Nowadays, more than 400
DIBs have been identified in the ISM (e.g., Hobbs et al. 2008; Geballe et al.
2011). No DIB carrier has been convincingly identified, although more
recent studies suggest that the DIB carriers may be complex molecules containing
carbon (see e.g., Herbig 1995; Snow \& McCall 2006; Cox 2011). Polycyclic
aromatic hydrocarbons (PAHs; e.g., Salama et al. 1999), fullerenes (e.g., Foing
\& Ehrenfreund 1994; Herbig 2000; Iglesias-Groth 2007), and polyatomic organic
molecules (e.g., Maier et al. 2011) are among the proposed DIB carriers. In
particular, the fullerenes - DIB hypothesis may also explain the intense UV
absorption band at 217 nm as due to fullerene-based molecules such as
buckyonions (multishell fullerenes) (e.g., Iglesias-Groth 2004) and 
hydrogenated fullerenes (e.g., Cataldo \& Iglesias-Groth 2009).

The 9577 and 9632 \AA~DIBs observed in a few hot reddened stars lie near two
electronic transitions of the C$_{60}$ cation observed in rare gas matrices
(Foing \& Ehrenfreund 1994). However, the presence of fullerenes in
astrophysical environments has been a matter of debate until recently when
Spitzer observations have provided evidence for C$_{60}$ and C$_{70}$ fullerenes
from planetary nebulae (PNe; Cami et al. 2010; Garc\'{\i}a-Hern\'andez et al.
2010, 2011a, 2012a), reflection nebulae (Sellgren et al. 2010) and the least
H-deficient R Coronae Borealis (RCB) stars (Garc\'{\i}a-Hern\'andez et al.
2011b,c). None of these environments is highly hydrogen-deficient. Furthermore,
the recent detection of C$_{60}$ fullerenes in PNe with normal H-abundances
(Garc\'{\i}a-Hern\'andez et al. 2010, 2011a, 2012a) suggests that large
fullerenes may be formed as decomposition products of hydrogenated amorphous
carbon (HAC) dust and that fullerenes may be not so exotic and can form under
conditions that are common to essentially all solar-like stars at the end of
their lives. 

   \begin{figure*}
   \centering
   \includegraphics[angle=0,scale=.45]{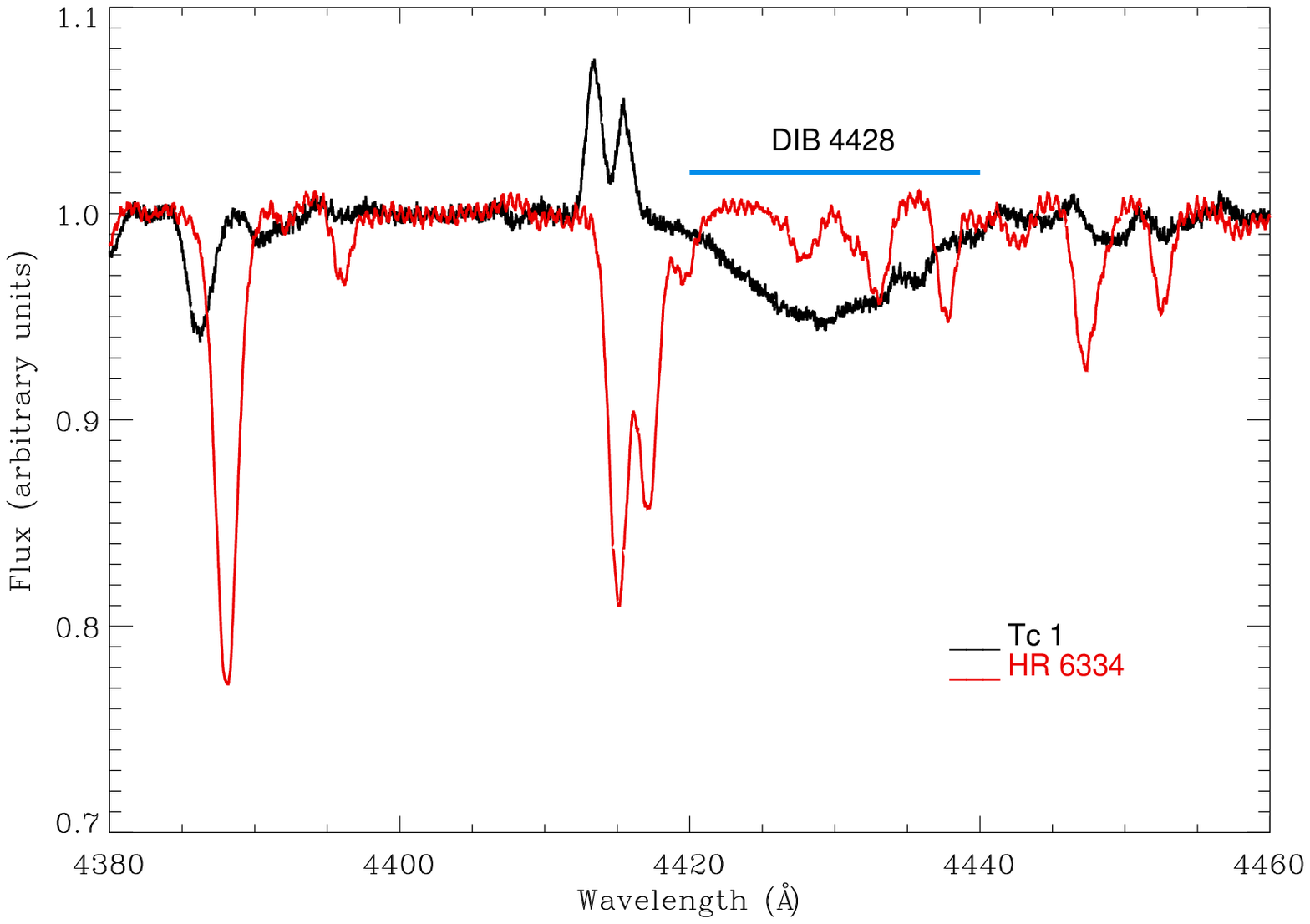}%
   \includegraphics[angle=0,scale=.45]{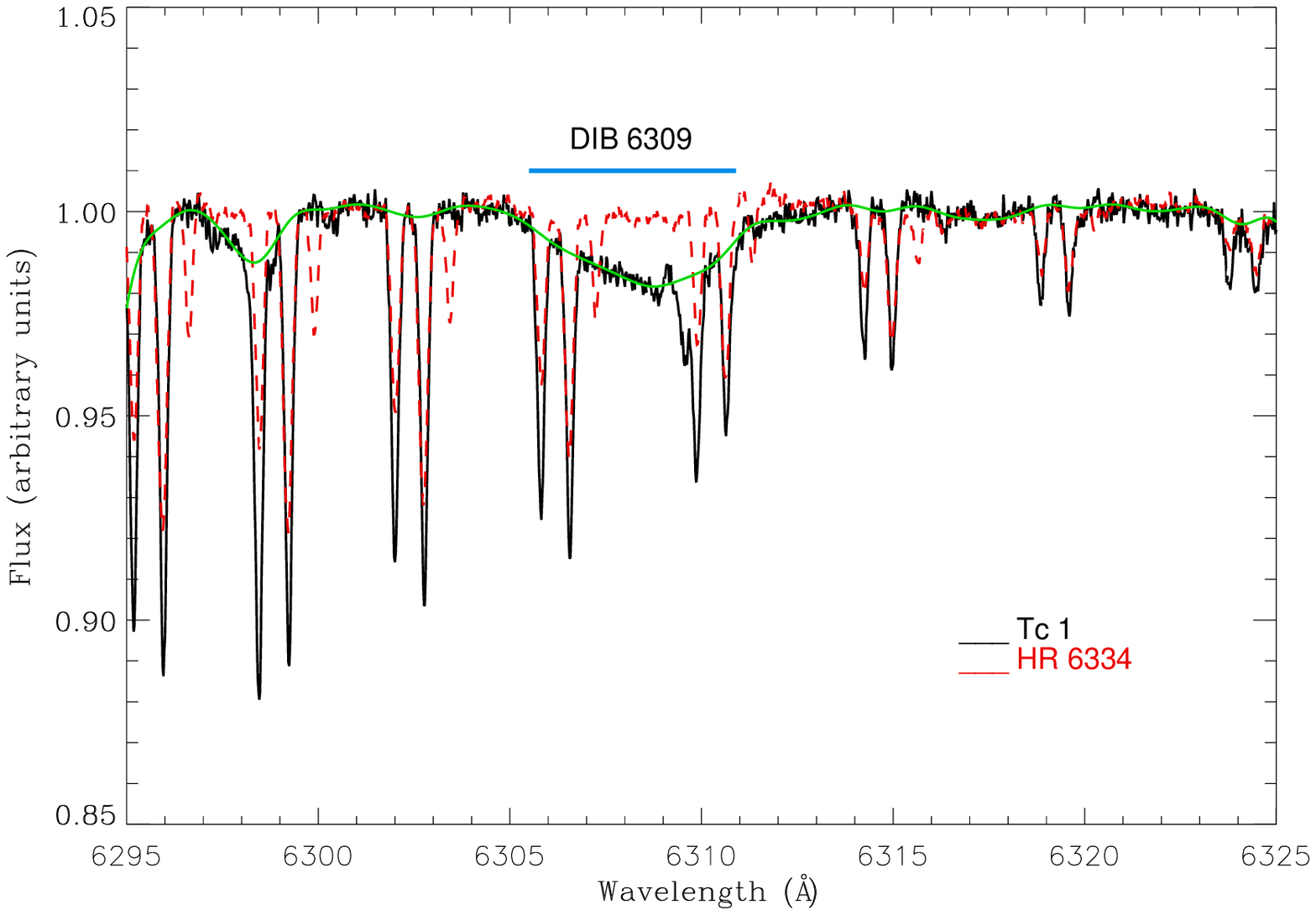}
    \caption{Spectra of Tc 1 (in black) and HR 6334 (in red) around the 
4428 \AA\ (left panel) and 6309 \AA\ DIBs (right panel). Both DIBs are found to
be unusually strong in Tc 1 while HR 6334 - with a higher E(B-V) of 0.42 - does
not show evidence of their presence. The telluric line corrected
spectrum of Tc 1 around 6309 \AA\ (right panel) is shown by the smooth line (in
green).}
              \label{Fig1}%
    \end{figure*}

Thus, fullerenes and related large carbon-based species (e.g., other fullerenes
as stable exohedral or endohedral metallo-complexes; Kroto \& Jura 1992) might
be ubiquitous in the ISM and continue to be serious candidates for DIB
carriers (e.g.,  Bettens \& Herbst 1996; Herbig 1995; Iglesias-Groth 2007).
However, a detailed analysis of the DIBs towards fullerene-containing -
accompanied or not by PAH molecules (e.g., Garc\'{\i}a-Hern\'andez et al. 2012b)
- astrophysical environments is mandatory before one can reach conclusions about
the nature of the DIB carriers. In this context, the recent infrared detection
of large fullerenes in PNe offers the beautiful opportunity for studying the
DIB spectrum of sources where fullerenes are abundant. In this {\it Letter} we
present for the first time a detailed inspection of the optical spectra of two
fullerene PNe, which permits us to directly explore the fullerenes - DIB
connection. 


\section{Optical VLT/UVES spectroscopy of fullerene PNe}

We acquired optical spectra of the fullerene PNe Tc 1 (B=11.1, E(B-V)=0.23;
Williams et al. 2008) and M 1-20 (B=13.7, E(B-V)=0.80; Wang \& Liu 2007). Tc 1
displays a fullerene-dominated spectrum with no clear signs of PAHs while M 1-20
also shows PAH features (Garc\'{\i}a-Hern\'andez et al. 2010). The observations
were carried out at the ESO VLT (Paranal, Chile) in service mode between May and
September 2011. The optical spectra were taken in the wavelength ranges
$\sim$3300-4500, 5750-7500, and 7700-9400 \AA~with UVES at the UT2 telescope
using the 2.4" slit with the standard setting DIC2 (390+760). This configuration
gives a resolving power of $\sim$15,000 and an adequate interorder separation.
We required R$\sim$15,000 to appropriately sample the relatively broad C$_{60}$
features together with the generally narrower DIBs (e.g., 5797, 5850, 6196, 6614
\AA). To detect weak broad ($\geq$4 \AA) C$_{60}$ features in the optical we
aimed for a minimum signal-to-noise (S/N) ratio of 200 around 3760 \AA. 

As comparison stars for Tc 1 we selected the nearby B-type star HR 6334 (B=5.1;
E(B-V)=0.42; Wegner 2003) and for M 1-20 HR 6716 (B=5.7; E(B-V)=0.22; Wegner
2003). Both comparison stars were observed on the same dates as the
corresponding PNe using the same VLT/UVES set-up. The observed spectra were
processed by the UVES data reduction pipeline (Ballester et al. 2000) and the
stellar continuum was fitted by using standard astronomical tasks in IRAF. 

    \begin{figure}
   \centering
   \includegraphics[angle=0,scale=.45]{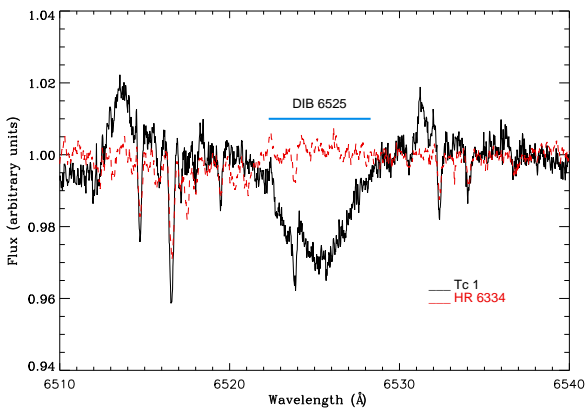}
    \caption{Spectral region around the new broad unidentified band at 6525 \AA\
in Tc 1 (in black) and HR 6334 (in red).}
              \label{Fig2}%
    \end{figure}

For Tc 1, we obtained 10-12 individual exposures (of 900 s each) in the
$\sim$3300-4500, 5750-7500, and 7700-9400 \AA~spectral ranges,  giving total
exposure times of 2.5-3 hours. The S/N in the continuum in the summed Tc 1
spectrum is $\sim$300 at 4000 \AA\ and higher than 350 at wavelengths longer
than 6000 \AA. For the fainter PN M 1-20, however, we only could obtain eight
individual exposures of 1800 s each in both $\sim$5750-7500 and 7700-9400
\AA~spectral regions, giving a total exposure time of four hours. The S/N in the
individual 1800 s exposures in the blue region (3300-4500 \AA) was too low
($<$10). During the execution of our service program we decided to use a binning
2x2 to increase the S/N around 4000 \AA\, but we only obtained two individual
exposures of 1800 s each. Thus, the S/N in the continuum in the best M 1-20
spectra is $\sim$20 at 4000 \AA\ and higher than 30 at wavelengths longer than
6000 \AA. Finally, an S/N in excess of 450 was easily achieved in the summed
spectra of the bright comparison stars HR 6334 and HR 6716 by using total
exposure times of several minutes.

Unfortunately, the S/N in the M 1-20 spectrum is too low to search for weak
and broad absorption bands (e.g., neutral C$_{60}$ features, see below) in its
spectrum but it was found to be enough to detect and characterize several of the
strongest and most common DIBs towards this PN with a rather high E(B-V) of
0.80.

\section{Electronic transitions of neutral C$_{60}$}

We have searched the high-quality (S/N$>$300) spectra of the PN Tc 1 (see above)
for electronic transitions of C$_{60}$.\footnote{C$_{70}$ is ten times less
abundant than C$_{60}$ (e.g., Garc\'{\i}a-Hern\'andez et al. 2012a) and we do
not find evidence for extra absorption around ($\pm$10 \AA) the expected
C$_{70}$ electronic transitions (e.g., Ajie et al. 1990).} The strongest
electronically allowed transitions of neutral gas-phase C$_{60}$ molecules, as
measured in the laboratory, are located around $\sim$3760, 3980, and 4024
\AA~with widths of 8, 6, and 4 \AA (Sassara et al. 2001; see also
Garc\'{\i}a-Hern\'andez et al. 2012b). The strongest C$_{60}$ transition seen in
the laboratory spectra is that at 3980 \AA~(Sassara et al. 2001).

Surprisingly, we can find no evidence for the presence of neutral C$_{60}$ in 
absorption (or emission) at the wavelengths of the expected electronic
transitions mentioned above. This is shown in Fig. A1 (in the appendix) where we
compare the Tc 1 velocity-corrected spectra\footnote{We applied average velocity
corrections of -105.8 kms$^{-1}$ ($\sim$-1.4 \AA) for Tc 1 and of 20.1
kms$^{-1}$ ($\sim$0.3 \AA) for HR 6334.} with those of the nearby B-type star HR
6334 around the expected positions of the strongest C$_{60}$ transitions. It is
to be noted here that the 3760 and 4024 \AA~bands coincide with several strong O
lines and He I 4026 \AA\  line, respectively, which hampers the
identification of broad and weak absorption features. However, the spectral
region around 3980 \AA~(Fig. A1 in the appendix, right panel) is free from any
contaminant and there is no evidence for the presence of the neutral C$_{60}$
feature at this wavelength. 

The one-sigma detection limits on the equivalent widths (EQWs) derived from our
Tc 1 spectra are 63, 20, and 12 m\AA~for the 3760, 3980, and 4024
\AA~C$_{60}$ transitions\footnote{One-sigma detection limits for the EQWs in our
spectra scale as $\sim$ 1.064  x FWHM / (S/N) (see e.g., Hobbs et al. 2008)
but this value for the 3760\AA~band is estimated by modeling and subtracting
the oxygen lines around 3760 \AA.}. This translates into column
densities of 6 $\times$ 10$^{12}$, 1 $\times$ 10$^{13}$, and 6 $\times$
10$^{12}$ cm$^{-2}$. We can compare this column density limit of
about 10$^{13}$ cm$^{-2}$ with estimates of the circumstellar density of
C$_{60}$ molecules, taking into account the total number of C$_{60}$ molecules
(N(C$_{60}$) = 1.8 $\times$ 10$^{47}$ for d= 2 kpc; Garc\'{\i}a-Hern\'andez et
al. 2011a) calculated from the IR C$_{60}$ features. By following
Garc\'{\i}a-Hern\'andez et al. (2012b) (Equations 1, 2, and 3), we can estimate
the density of C$_{60}$ molecules by assuming a spherical shell of radius
(R$_{out}$ $-$ R$_{in}$) and a uniform number of C$_{60}$ molecules throughout
the shell. Thus, considering that L=1480 L$_{\odot}$ for Tc 1 (Pottasch et al.
2011), we obtain dust temperatures (T$_{d}$) of 415 K at $\sim$18 au and 100 K
at 301 au, where the first temperature is the C$_{60}$ excitation temperature
(Garc\'{\i}a-Hern\'andez et al. 2011a) and the latter temperature corresponds to
the minimum temperature of the dust to be detected in the mid-IR by Spitzer. For
a distance of 2 kpc, we estimate a circumstellar density of fullerenes
n(C$_{60}$)=0.46 cm$^{-3}$ and a C$_{60}$ column density along the path
(R$_{out}$ $-$ R$_{in}$) of $\sim$2 $\times$ 10$^{15}$ cm$^{-2}$; a lower
C$_{60}$ column density of 3 $\times$ 10$^{14}$ cm$^{-2}$ is obtained for
L=10$^{4}$ L$_{\odot}$. Thus, our estimate of the C$_{60}$ column density is
200$-$300 times higher than the observed upper limits. It is to be noted here
that although the Spitzer/IRS observations contain a marginal amount of spatial
information ($\sim$2"/pixel) and that mid-IR images at much higher spatial
resolution would be desirable, Bernard-Salas et al. (2012) presented tentative
evidence that the 8.5$\mu$m emission (and attributed to C$_{60}$) in Tc 1 is
extended and peaks at 2-3 pixels ($\sim$6400-9700 au for d= 2 kpc) from
the central star. By assuming that fullerenes are uniformly distributed in a
shell of R$_{out}$=9700 au and R$_{in}$=6400 au, the estimated C$_{60}$
column density ($\sim$1 $\times$ 10$^{12}$ cm$^{-2}$) is  a factor between 6 and
10 below our observed upper limits.

In short, our C$_{60}$ column density values estimated from the C$_{60}$ IR
bands could explain the non-detection of the electronic C$_{60}$ transitions in
our Tc 1's optical spectra only if the C$_{60}$ emission peaks far away
from the central star. However, we cannot discard that the line of sight to Tc
1 may not intersect the fullerene-rich regions of the circumstellar shell (e.g.,
if the fullerenes may be formed in clumps). On the other hand, one can
speculate that the strongest 3980\AA~C$_{60}$ band - perhaps the other
electronic bands, too - could be suppresed  if the fullerenes are in the
solid-state phase (e.g., Evans et al. 2012; Garc\'{\i}a-Hern\'andez et al.
2012a). However, a laboratory spectrum of solid-state C$_{60}$ in n-hexane
displays the same transitions seen in the gas-phase (F. Cataldo, private
communication). An alternative and more exotic explanation may be that the
mid-IR features seen in Tc 1 are not due to C$_{60}$ and C$_{70}$ solely, being
contaminated by other more complex fullerene-based molecules. This latter
interpretation seems to be supported by our study of the DIBs towards Tc 1 and
presented below. 

\section{Diffuse interstellar bands in fullerene PNe}

We have followed the catalog of DIBs measured in the high-S/N HD 204287's
spectrum (Hobbs et al. 2008) to search them in the VLT/UVES spectrum of Tc 1 and
M 1-20. However, we concentrate here on analyzing eight of the strongest DIBs
most commonly found in the ISM as well as on detecting unusually strong
DIBs (i.e., not present in the nearby comparison stars and/or in Hobbs et al.
2008), which could be potentially due to fullerenes or fullerene-based
molecules. Thus, we can compare the characteristic of most common DIBs for both
PNe in our sample as well as with existing literature data on field-reddened
stars (e.g., Luna et al. 2008). The exhaustive analysis of the weaker DIBs
listed by Hobbs et al. (2008) and also detected in Tc 1's spectrum will be
published elsewhere.

Our list of DIBs in Tc 1 and M 1-20 are listed in Table A1 (in the appendix),
where we give the measured central wavelength, FWHM, EQW, the S/N in the
neighboring continuum, and the normalized equivalent widths EQW/E(B-V). For
comparison we also list the EQW/E(B-V) measured in HD 204827 and field-reddened
stars by Hobbs et al. (2008) and Luna et al. (2008). It should be noted that we
could not estimate the total absorption of the 6993 and 7223 \AA~DIBs (not shown
in Table A1 in the appendix) in our PNe because of the strong meddling from the
telluric lines. For the well-studied DIBs at 5780, 5797, 5850, 6196, 6270, 6284,
6380, and 6614 \AA\footnote{Note that a possible exception is the 6284\AA~DIB
but its strength is known to be not very well correlated with the interstellar
reddening (e.g., Luna et al. 2008).}, the EQW/E(B-V) in fullerene PNe agree
reasonably well with the values reported in HD 204827 (Hobbs et al. 2008) and
field-reddened stars (Luna et al. 2008). The observed nearby comparison stars HR
6334 and HR 6716 also display EQW/E(B-V) values completely consistent (within
the errors) with those measured in our PNe. This indicates that the carriers of
the latter well-studied DIBs are not particularly overabundant in fullerene PNe.

Interestingly, the well-studied DIB at 4428 \AA\ as well as the weaker 6309 \AA\
DIB listed by Hobbs et al. (2008) are found to be unusually strong towards Tc 1.
These last DIBs are not detected in the nearby comparison star HR 6334 with a
higher E(B-V) of 0.42. Fig. 1 compares the 4428 and 6309 \AA\ DIBs in Tc 1 with
those in HR 6334. Adopting a Lorentzian profile for the 4428 \AA\ DIB (Snow et
al. 2002), we obtain EQW=860 m\AA, which is at least a factor of two greater
than expected for the low reddening of E(B-V)=0.23 in Tc 1; see e.g., Fig. 6 and
15 in Snow et al. (2002) and van Loon et al. (2012). We note that there is
tentative evidence for an unusually strong (5-10\% of the continuum and
EQW=2579$\pm$786 m\AA) 4428 \AA\ DIB in M 1-20, too  (see Fig. A2 in the
appendix), something that supports our finding in Tc 1. However, we prefer to be
cautious until higher S/N M 1-20 spectra are obtained. On the other hand, the
uncommon 6309 \AA\ DIB in Tc 1 appears to be three times more intense than that
observed in HD 204287 by Hobbs et al. (2008) (see Table A1 in the appendix).
Finally, an unidentified broad feature at 6525 \AA\ (FWHM $\sim$5 \AA, EQW=173
m\AA) is detected in the Tc 1 spectrum (Fig. 2). Fig. 2 shows that the 6525 \AA\
band is real because it is not detected in the spectrum of the comparison star
HR 6334 taken with the same UVES setup and at the same time. 

\section{Fullerenes - DIB connection}

Our finding of an unusually strong 4428 \AA\ DIB towards Tc 1 (see left panel of
Fig. 1) necessarily prompts the idea that the 4428\AA\ DIB carrier may be
related with fullerenes or fullerene-based molecules (Iglesias-Groth 2007).
Remarkably, photo-absorption theoretical models of several large fullerenes such
as C$_{80}$, C$_{240}$, C$_{320}$, and C$_{540}$ predict their strongest
transitions very close in wavelength ($\pm$10 \AA) to this 4428 \AA\ DIB
(Iglesias-Groth 2007)\footnote{The same models do not predict a 4428 \AA\ band
for the C$_{60}$ fullerene, in agreement with fullerene laboratory spectroscopy
(e.g., Sassara et al. 2001).}. The theoretical spectra of several multishell
fullerenes (buckyonions such as C$_{60}$@C$_{240}$ and
C$_{60}$@C$_{240}$@C$_{540}$) reported by Iglesias-Groth (2007) also display a
strong 4428 \AA\ band. In this context, the broad 4428 \AA\ band may be well
explained by the superposition of the transitions of fullerenes bigger than
C$_{60}$ and buckyonions. This would be consistent with the recent exhaustive
study of the 4428\AA\ DIB by van Loon et al. (2012), which suggests the carrier
to be a large, compact, and neutral molecule that is relatively resistant to
impacting energetic photons or particles. 

Another interesting feature is that our Tc 1 spectra also lack  the unidentified
4000 \AA~band (see Fig. A1 in the appendix) that is detected in the RCB star DY
Cen (Garc\'{\i}a-Hern\'andez et al. 2012b). Garc\'{\i}a-Hern\'andez et al.
(2012b) suggested that the mid-IR features at $\sim$7.0, 8.5, 17.4, and 18.8
$\mu$m and the unidentified 4000 \AA\ band in DY Cen are likely due to
proto-fullerenes (PFs) or fullerene precursors rather than to C$_{60}$.
Interestingly, DY Cen displays a `normal' 4428\AA\ DIB, supporting  the claim
that fullerenes and fullerene-based molecules such as buckyonions are not
especially overabundant toward DY Cen. Thus, the unusually strong 4428\AA\ DIB
and the lack of the unidentified 4000 \AA~band in Tc 1 may indicate an efficient
conversion of PFs to fullerenes and fullerene-based molecules in its
circumstellar envelope (see e.g., Duley \& Hu 2012).

Furthermore, the apparent lack of the strongest electronic transitions of the
C$_{60}$ molecule in Tc 1 may indicate that we are not seeing emission from
isolated, free C$_{60}$ molecules, which would explain the variable
properties of the mid-IR C$_{60}$ spectral features observed in fullerene PNe
(Bernard-Salas et al. 2012; Garc\'{\i}a-Hern\'andez et al. 2012a). At present,
the HAC's photochemical processing is the most likely C$_{60}$ formation route
in the complex circumstellar envelopes of PNe (Garc\'{\i}a-Hern\'andez et al.
2012a; Bernard- Salas et al. 2012; Micelotta et al. 2012). Larger fullerenes may
grow from pre-existing C$_{60}$ molecules (Dunk et al. 2012) that may be
supplied by the photochemical processing of HAC dust, opening the possibility of
forming other fullerene-based molecules such as buckyonions and fullerene
adducts. Indeed, fullerenes and PAHs may be mixed in the circumstellar envelopes
of fullerene PNe (e.g., M 1-20) and fullerene/PAH adducts may form via
Dies-Alder cycloaddition reactions (Garc\'{\i}a-Hern\'andez et al. 2013).

Fullerene clusters or fullerene-based molecules such as buckyonions,
fullerene/PAH adducts may still be excited by stochastic heating (e.g., from UV
photons) emitting through the same IR vibrational modes. Indeed, very recent
laboratory work demonstrates that fullerene/PAH adducts - such as
C$_{60}$/anthracene Diels-Alder adducts - display mid-IR features strikingly
coincident with those from C$_{60}$ and C$_{70}$ (Garc\'{\i}a-Hern\'andez et al.
2013). Unfortunately, the synthesis of multishell fullerenes (buckyonions) in
the laboratory is challenging because their insolubility does not permit us to
extract and separate these species from the carbon soot in which they are
present in small amounts (F. Cataldo, private communication).

In summary, we propose that the 4428 \AA\ DIB (possibly also the 6309\AA\ DIB
and the new 6525 \AA\ band) is probably related to fullerenes
bigger than C$_{60}$ (e.g., C$_{80}$, C$_{240}$, C$_{320}$, and C$_{540}$) and
buckyonions (e.g., C$_{60}$@C$_{240}$, C$_{60}$@C$_{240}$@C$_{540}$) in the Tc
1 circumstellar environment. This possible fullerenes - DIB connection was
previously suggested by Iglesias-Groth (2007) from theoretical considerations.

\begin{acknowledgements}
We acknowledge the referee Jacco van Loon for very useful comments that
helped to improve the paper. We also acknowledge A. Manchado for his help
during the early stage of this project and F. Cataldo for providing us
with his fullerene laboratory data and expertise that helped to improve this
Letter. This work is based on observations obtained with ESO/VLT under the
program 087.D-0189(A). D.A.G.H. acknowledges support  provided by the Spanish
Ministry of Economy and Competitiveness under grant AYA$-$2011$-$27754.  
\end{acknowledgements}

\Online
\begin{appendix}
\section{Figures A1 and A2 and Table A1}

  \begin{figure*}
   \centering
   \includegraphics[angle=0,scale=.45]{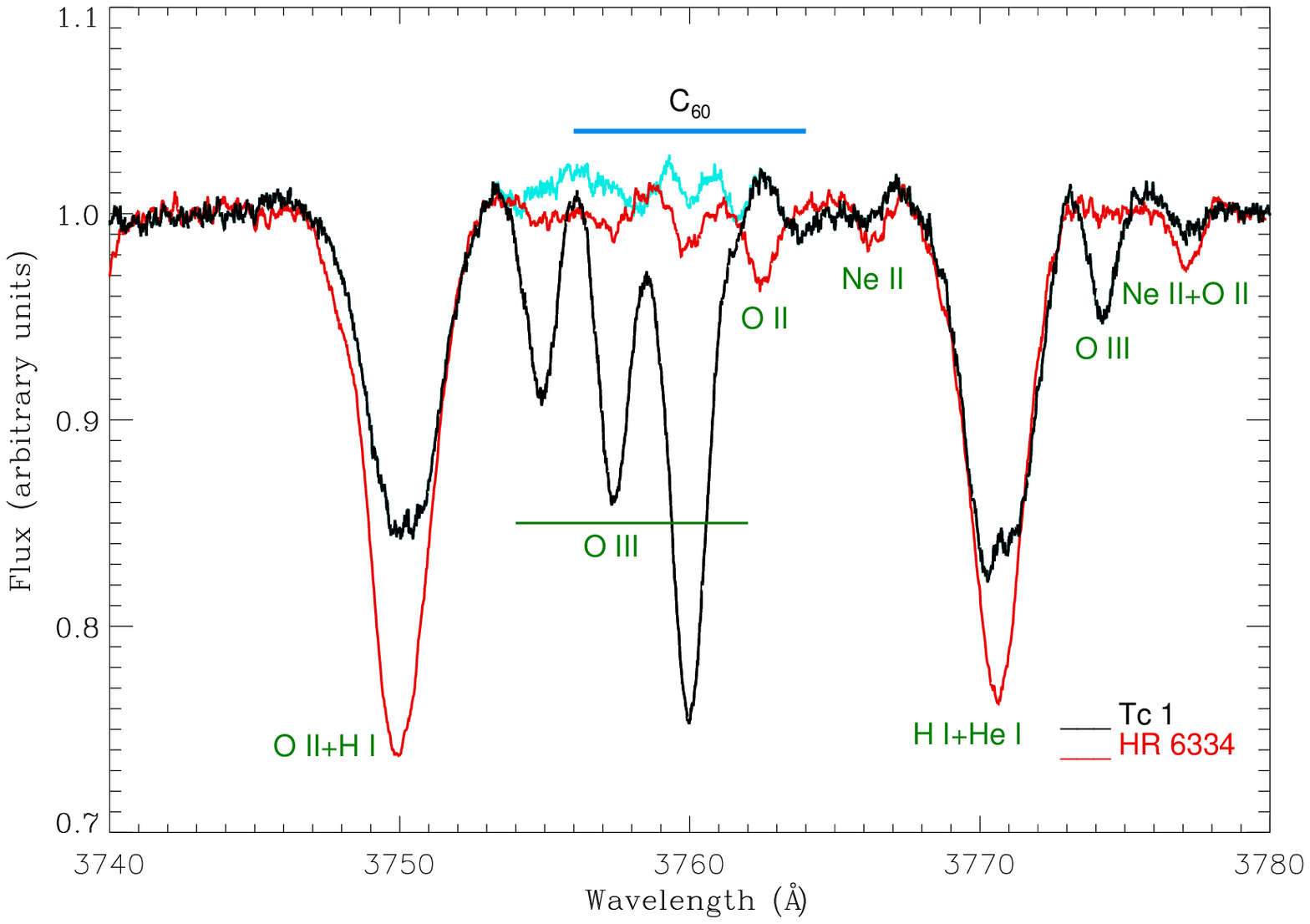}%
   \includegraphics[angle=0,scale=.45]{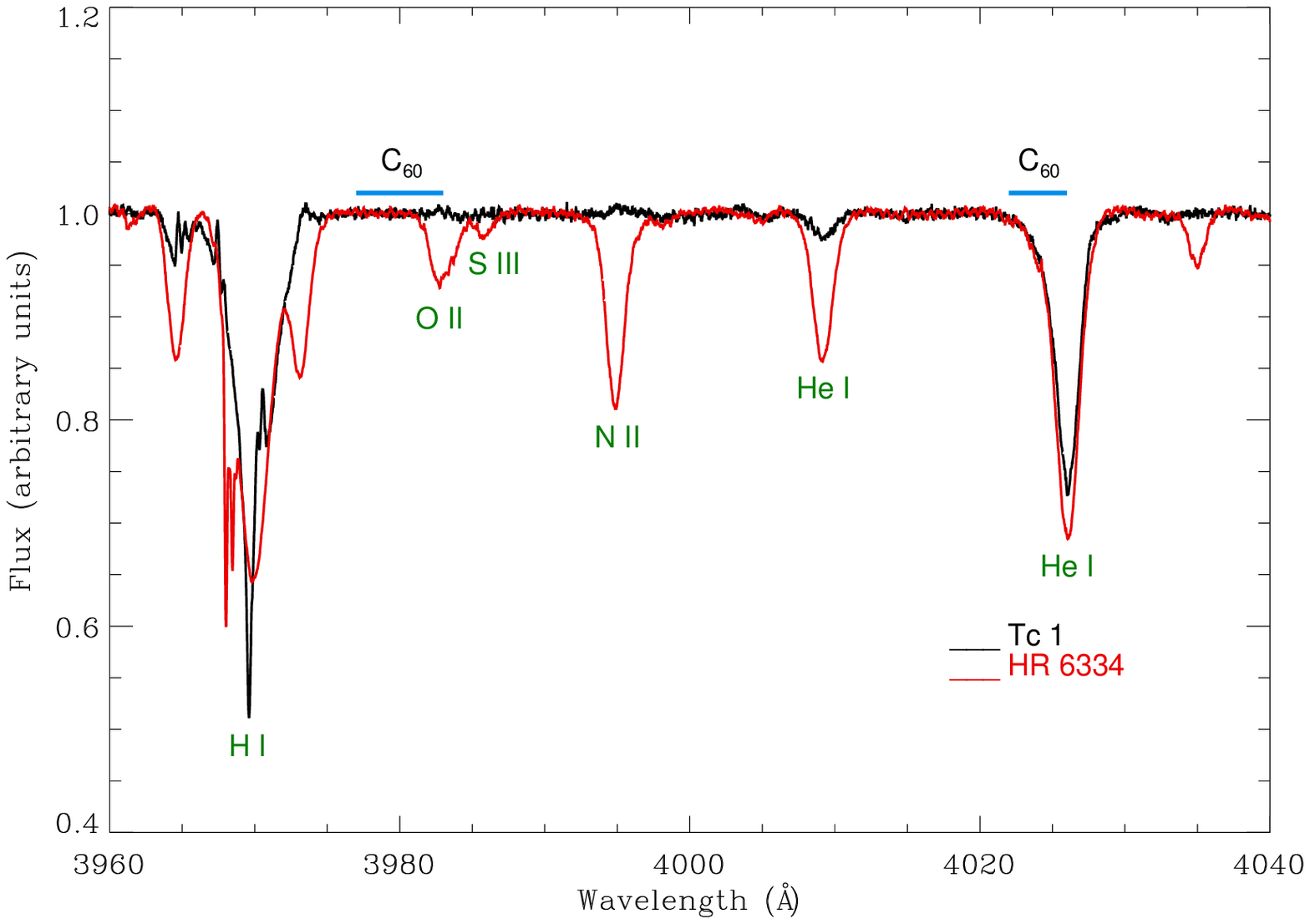}
    \caption{Velocity-corrected spectra of Tc 1 (in black) and HR 6334 (in red)
around 3760 \AA\ (left panel) and 4000 \AA\ (right panel) where the atomic line
identifications are indicated (in green). The expected positions (and FWHMs) of
the C$_{60}$ features are marked on top of the spectra. There is no
evidence (additional absorption) in Tc 1 for the neutral
C$_{60}$ features at 3760, 3980, and 4024 \AA. The residual
spectrum (in cyan) obtained by subtracting the oxygen lines around 3760 \AA~is
also shown.}
\label{FigA1}%
    \end{figure*}

   \begin{figure*}
   \centering   
   \includegraphics[angle=0,scale=.45]{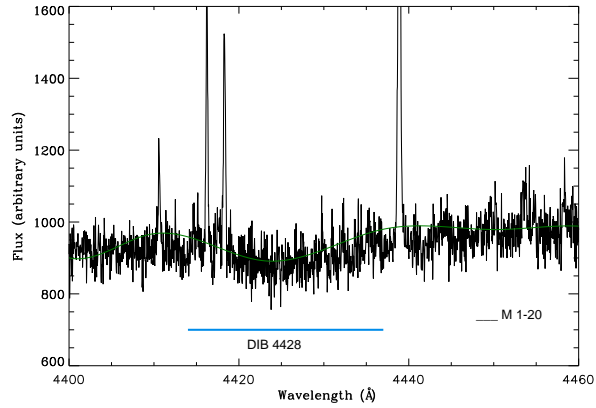}
    \caption{Spectrum of M 1-20 (in black) around the broad 4428 \AA\
DIB. A heavily smoothed and emission line corrected spectrum is also
overplotted (in green).} \label{FigA2}%
    \end{figure*}

\begin{table*}
\tiny
\caption{\label{t1}Diffuse interstellar bands in fullerene PNe.\tablefootmark{a}}
\centering
\begin{tabular}{lccccccccccc}
\hline\hline
Tc 1            &      &     &     &            &M 1-20           &      &     &     &            &   Hobbs et al.         &     Luna et al.      \\
$\lambda$$_{c}$ & FWHM & EQW & S/N & EQW/E$_{B-V}$ & $\lambda$$_{c}$ & FWHM & EQW & S/N & EQW/E$_{B-V}$ & EQW/E$_{B-V}$ & EQW/E$_{B-V}$\\
($\AA$)         &($\AA$) & (m$\AA$) & &  ($\AA$/mag)       &($\AA$) & ($\AA$) & (m$\AA$)  &     &  ($\AA$/mag)          &   ($\AA$/mag)                     &      ($\AA$/mag)                \\
\hline
4428.10\tablefootmark{b} & 19.35 & 860 &329 &3.74    & 4426.56\tablefootmark{b}    & 19.94\tablefootmark{c}    & 2579\tablefootmark{c}   & 20\tablefootmark{d}  &  3.22  &  1.10    &  $\dots$  \\
5780.40 & 2.06  & 105 &416 &0.46    & 5780.66   & 2.17      & 359       & 32       & 0.45      &  0.23    &  0.46     \\
5796.88 & 0.99  &  42 &357 &0.18    & 5797.31   & 1.14      & 155       & 53       & 0.19      &  0.18    &  0.17     \\
5849.60 & 1.35  &  11 &346 &0.047   & 5850.02   & 1.23      &  70       & 37       & 0.087     &  0.086   &  0.061    \\
6195.85 & 1.17  &  13 &549 &0.056   & 6196.18   & 0.95      &  36       & 68       & 0.045     &  0.034   &  0.053    \\
6269.77 & 1.93  &   9 &548 &0.037   & 6270.19   & 2.48      & 130       & 66       & 0.16      &  0.069   &  $\dots$  \\
6283.93 & 4.91  & 316 &579 &1.38    & 6283.77   & 5.25      & 706       & 61       & 0.88      &  0.41    &  0.90     \\
6308.90 & 2.98  &  42 &530 &0.18    & $\dots$   & $\dots$   & $\dots$   & $\dots$  & $\dots$   &  0.049   &  $\dots$  \\
6379.08 & 1.39  &  12 &513 &0.050   & 6379.56   & 1.30      &  93       & 56	   & 0.12      &  0.085   &  0.088    \\
6525.15 & 4.79  & 173 &522 &0.75    & $\dots$   & $\dots$   & $\dots$   & $\dots$  & $\dots$   &  $\dots$ &  $\dots$  \\
6613.50 & 1.42  &  36 &412 &0.16    & 6613.78   & 1.23      & 167       & 77       & 0.21      &  0.149   &  0.21     \\
\hline
\end{tabular}
\tablefoot{
\\
\tablefoottext{a}{The 3-$\sigma$ errors in the EQWs scale like $\sim$3 $\times$
FWHM/(S/N) while we estimate that the FHWMs in Tc 1 are precise to the 0.03 \AA\
level (less for M 1-20).}
\\
\tablefoottext{b}{The characteristics of this DIB are estimated by adopting a
Lorentzian profile (see e.g., Snow et al. 1992).}
\\
\tablefoottext{c}{Best estimates found by clipping out the narrow emission lines and smoothing the spectrum
with boxcar 15. The error in the quoted EQW is estimated to be $\sim$786 m\AA.}
\\
\tablefoottext{d}{S/N in the original spectrum.}
}
\end{table*}

\end{appendix}

\end{document}